\documentclass{mem}
\usepackage{txfonts}
\usepackage{balance}
\usepackage{graphicx}
\usepackage[a4paper]{hyperref}
\idline{75}{282}

\begin{document}

\title{
Radio measurements of constant variation, and perspectives with ALMA
}
   \subtitle{}
\author{
Fran\c{c}oise \,Combes
          }

  \offprints{F. Combes}
 
\institute{
Observatoire de Paris, LERMA,
61 Av. de l'Observatoire, 
F-75014, Paris, France \\
\email{francoise.combes@obspm.fr}
}

\authorrunning{Combes }

\titlerunning{Future of radio measurements}

\abstract{
The present constraints on fundamental constant variation ($\alpha$ and $\mu$)
obtained in the radio range are reviewed, coming essentially from absorption
lines in front of quasars of intermediate to high-z galaxies,
through CO, HI, OH, HCO$^+$, HCN .. lines up to NH$_3$ and CII.
 With ALMA, the sensitivity to detect radio continuum sources
in narrow bands will increase by an order of magnitude,
and the expected progress is quantified. The relative advantage of the radio domain with respect to the optical one is emphasized.
}
\maketitle{}

\section{Introduction}
 Molecular absorption lines in the radio domain are very narrow,
and the high spectral resolution is well suited to the study
of variation of constants, with space and time. 
Observations of absorption lines in the millimeter domain
 have been developped in the 1990's in both our own Milky
Way (Lucas \& Liszt 1994) and in very distant galaxies (Wiklind \& Combes
 1994).
Up to now,  at intermediate and high redshift, 
between z$=$0.25 to z$=$0.89, only four
absorption lines systems have been detected in the millimeter range
and a fifth system at 0.765, at the OH-18cm lines (Kanekar et al 2005). 
Out of these 5 systems, 3 are intervening lensing galaxies (and the 
background quasar is multiply imaged), and 2 correspond to an absorption
of the host (PKS1413+135, B3-1504+377).
Provided that the background continuum
source is strong enough, a large
number of different molecules and rotational transitions can been
observed (for an overview see Combes \& Wiklind 1996; Wiklind \& Combes
1994, 1995, 1996a,b, 1997).
The detectability depends mainly on the background source intensity, 
and with the increased sensitivity of ALMA, we could expect to
detect 30-100 times more sources.

\section{Space variation at z=0} 
Even at z=0, molecular emission and absorption lines bring very stringent constraints
on the variation of constants with space. In the current quintessence models,
proposed to solve the dark energy problem, hypothetical scalar fields couple
to the ordinary matter, leading to long-range forces and variations of the fundamental constants 
(e.g. Garcia-Berro et al 2007). In particular, the 
Chameleon theory predicts different coupling constants, and different baryon masses
between low and high densities, or for different gravitational potentials.
 Since the difference in densities between
the Earth environment, and molecular clouds can be as large
as 20 orders of magnitude, the Milky Way itself is an ideal laboratory
to test such variations. 
One of the abundant molecules, which has the largest sensitivity to the 
electron to proton mass ratio  $\mu =m_e/m_p$, is the ammoniac, NH$_3$,
and its peculiar inversion lines (Flambaum \& Kozlov 2007). It can be shown that
the position of the lines are $\Delta v/c= K_i\, \Delta\mu/\mu$, with 
$K_i$ = 1 for rotation lines, and $K_i$= 4.46 for NH$_3$, therefore comparing the two give
$(z_{inv}-z_{rot})/(1+z) =3.46\, \Delta\mu/\mu$.

Compared to rotational lines of CCS, HC$_3$N, with different sensitivity to  $\mu$,
it was possible to constrain 
the variation  $\Delta \mu/\mu$ at the  $10^{-8}$ level. There might remain
systematic kinematical variations, 
differences of $\Delta$V = 10 m/s corresponding to $\Delta \mu/\mu =\, 10^{-8}$
(Levshakov et al 2008, Molaro et al 2009).

\section{High redshift absorption}

Then to probe time variations, a large number of high redshift systems are required.
 However, these are quite rare in molecular absorptions, since dense molecular
clouds are found only in the disk of spiral galaxies, and the cross-section 
is small in front of background quasars. The 
first system was found towards the BL Lac object PKS1413+135 (Wilkind \& Combes 94). 
It is a self-absorbing system, where HI-21cm absorption was already detected,
pointing to high column densities.
 When an intervening galaxy is on the line of sight towards
a quasar, the probability of lensing is high, and the amplification
of the quasar flux is a selection bias. This is therefore not a surprise that
three out of the five known systems are gravitationally lensed objects:
PKS1830-211, B0218+357 and PMN J0134-0931. 
 Due to the special geometry for optimum
amplification, intervening redshifts range up to z$\sim$1,
while the quasar is at z$\sim$2. It becomes difficult to find higher
redshift quasars, that are strong enough in the millimeter, since
most radio sources have a synchrotron steep spectrum. The cosmological 
K-correction works against us in this domain. 
A solution is to follow the higher redshift continuum sources to the
centimetric domain. 
Let us briefly describe the specific features of the absorbing systems.

\subsection{PKS1413+135 z=0.247}
This is the narrowest absorbing system, with
two velocity components, each narrower than
1 km/s. It is a BL Lac, very variable continuum source,  since
the radio jets are seen almost along the line of sight. This small
angle, with a large Lorentz factor of the jets, produces superluminal motions.
 The absorbing galaxy appears edge-on, and the absorbing molecular
cloud is  CO
optically thick, with N(H$_2$) $>\, 10^{22}$ cm$^{-2}$, and an 
visual extinction of A$_v >$ 30 mag. The comparison of the position of the HI-21cm
line and molecular lines constrains the variation of constants (Wiklind \& Combes 1997).

\subsection{B3 1504+377 z=0.672}
This is the second self-absorbing system, with a narrow component
(may be corresponding to the outer disk), and a broad component, at 330km/s
(may be widened by non-circular motions in the nucleus).
Up to 7 different molecular lines have been detected,  
and the chemistry of the absorbing diffuse gas is unexpected;
with high HCO$^+$ abundance.  The comparison of the HI-21cm position
with the molecular lines reveals strong differences in velocity,
which prevent to use this system for the constant variation study.

\subsection{B0218+357  z=0.685}
This system corresponds to the largest column density 
 N(H$_2$) $\sim \,10^{24}$ cm$^{-2}$;
all three CO isotopes (CO, $^{13}$CO and C$^{18}$O)
 have optically thick lines, and this allowed the detection of
many molecules (Combes \& Wiklind 1997). It was for instance possible to search
for molecular oxygen O$_2$ without atmospheric absorption,
with the redshifted lines at 368 and 424 GHz: the abundance
ratio was constrained to O$_2$/CO $<$ 2 10$^{-3}$,
meaning that most of the oxygen should be in atomic form 
(Combes \& Wiklind 1995, Combes et al 1997).

The system is a gravitational lens, with two
images of the quasar, and an Einstein ring. 
The image separation is 335mas (1.8kpc),
well determined through radio interferometry
(VLA, MERLIN, Browne et al 1993).
Substructures have been resolved in each
of the two images, in the form of an
extended radio jet component and two
 bright cores, through VLBI at 8.4 GHz, with 1mas 
synthetized beam (Biggs et al 2003).
The lens is a spiral galaxy almost face-on,
and spiral arms complicate the lens
analysis, to obtain the Hubble constant value from
the time delay (York et al 2005).

\subsection{PKS1830-211 z=0.88582}
This system is also a gravitational lens with two images and an Einstein ring.
The redshift of the lens was discovered in the millimeter range, through
sweeping of the band with the SEST telescope (after 14 GHz sweep in 14 tunings, already 2 lines were detected, Wiklind \& Combes 1996a, 1998). The two images were resolved with OVRO by Frye et al (1997). There are two velocity components, each one absorbing in front a different quasar image. This allows to monitor the variations and the time delay between the two images with a single dish only, 
without resolving the 2 images. The monitoring during 3 years 
(1h per week) determined a delay of 24$\pm$5 days, yielding a value for the Hubble constant of H$_0 = 69 \pm 12$ km/s/Mpc. An isotopic survey of the various molecules revealed different abundances than at z=0, due to the predominance of young stars, given the relative youth of the galaxy (Muller et al 2006).

\subsection{PMN J0134-0931 z=0.7645}
The z=0.7645 lens produces 6 images of the background quasar at z=2.22. HI-21cm and OH-18cm lines have been detected at GBT (Kanekar et al 2005). The main OH lines at 1665 and 1667 MHz are in absorption, while the satellite line at 1720 MHz is in emission. The other satellite line at 1612 MHz is in absorption, such that the sum of the two satellite lines is compatible with noise. The two satellite lines are therefore conjugate, which comes from the peculiar excitation of the upper levels through far-infrared radiation. This conjugate character ensures that the gas comes from the same regions, and therefore intrinsic velocity offsets will not perturb the constant variation study.

\section{Variations of constants}
 Constraining results have been obtained first in the optical domain, through absorption lines in front of quasars, with the Alkali Doublet (AD) method (lines of CIV, SiII, SiIV, MgII, AlIII...), and then the "many multiplet" method which takes advantage of a greater statistics (see e.g. the review by Petitjean et al 2009). A possible variation of the fine structure constant $\Delta\alpha/\alpha = (-0.6 \pm 0.15)\, 10^{-5}$ was reported (Murphy et al 2003) over redshifts between 0.2 and 3.7. However, with the same method at VLT, Srianand et al (2004) and Chand et al (2006) do not find any variation, with 
$\Delta\alpha/\alpha = (-0.6 \pm 0.6)\, 10^{-6}$ over a comparable redshift range. It is therefore interesting to compare with other methods.

\subsection{Constant variation with radio absorption}

According to the lines and molecules considered, it is possible to have constraints on $\alpha$, $\mu= m_e/m_p$, the gyromagnetic constant of the proton $g_p$, or a combination of them, 
such as $y= \alpha^2 g_p \mu$. The HI 21cm hyperfine transition is proportional to $\mu_p \mu_B / h a^3$, with the Bohr radius $a = h^2/m_e e^2$, and $\mu_p = g_p e h /4 m_p c$, $\mu_B = e h/2 m_e c$. On their side, the CO rotation lines frequencies vary as $ h / (M a^2)$, with  $M$ the reduced mass of the molecule. 
The great advantages of the radio domain are the large spectral resolution, up to $R=10^6$ thanks to the  heterodyne techniques, and the very narrow lines due to cold gas on the line of sight.

\subsection{Variation of y, $\alpha$ and $\mu$}

 A global comparison of all molecular lines observed with the HI-21cm absorption lines in PKS1413 and B0218 systems, the two narrowest line systems, have given quite stringent constraints,
$\Delta y/y = (-0.20 \pm 0.44)\,  10^{-5}$  for PKS1413 and
$\Delta y/y = (-0.16 \pm 0.54)\,  10^{-5}$  for B0218 (Murphy et al 2001).
The  precision is comparable to the MM method, with a limited number of absorbing systems. The radio method would then be very powerful with more systems detected. 
The main problem with this method is a possible kinematical bias, i. e. that the HI absorbing medium is not at the same velocity as the CO absorbing one. Of course, it is an issue only if a positive variation is detected. Since the velocity offset is expected to be randomly distributed, the solution is to observe a large number of systems. The precision will then be only limited by the dispersion in the offsets. This can be measured on the large number of HI and HCO$^+$ absorbing lines detected towards quasars in the Milky Way (Lucas \& Liszt 1998). The
dispersion is of only 1.2 km/s, corresponding to $\Delta y/y = 0.4 \, 10^{-5}$. 

In the special system PMN0134, although no HCO$^+$ nor other molecules have been detected with IRAM, ATCA or VLA, the use of the OH lines, together with HI, have also brought interesting constraints, less dependent on a possible velocity offset, because of the conjugate emission/absorption of the satellite OH lines.
Together with B0218, the value of  $F = g_p [\alpha^2/\mu]^{1.57}$ has been found not to vary, i.e.  
$\Delta F/F = (0.44 \pm 0.36^{stat} \pm 1.0^{syst})\, 10^{-5}$  for $0 < z  < 0.7$ (Kanekar et al 2005). This is a null variation with a 
sensitivity at 2$\sigma$  of $\Delta\alpha/\alpha \sim 6.7\, 10^{-6}$ and
$\Delta\mu/\mu \sim 1.4\, 10^{-5}$ over half of the age of the universe.

\subsection{Radio 21cm and UV combined}

Another method has combined several systems observed both in HI-21cm absorption and in UV, to constrain the parameter $y= \alpha^2 g_p \mu$. With 9 absorbing systems, over a redshift range $0.23 < z < 2.35$, null evolution was found, with
$\Delta y/y = (0.63 \pm 1)\, 10^{-5}$ or $dy/dt/y = (0.6 \pm 1.2)\, 10^{-15} /yr$ (Tzanavaris et al 2007). The precision is clearly limited by intrinsic
line of sight velocity differences, with random sign, of amplitude $\sim$ 6 km/s.
Combining with $\alpha$ variation constraints, and restricting at high redshift systems, this gives a constraint in $\mu$ of $\Delta\mu/\mu = (0.58 \pm 1.95)\, 10^{-5}$, which is in contradiction with the results from Reinhold et al (2006), who found with Lyman band of H$_2$ lines a positive detection of variation, $\mu$ being higher in the past.

\subsection{Radio centimeter lines}
 The high sensitivity if the NH$_3$ inversion lines to variation in the $\mu$ ratio was used by Henkel et al (2009) in a recent multi-line study of  PKS1830-211  at $z \sim 0.9$. 

They have observed with the Effelsberg telescope NH$_3$, HC$_3$N, SiO, C$^{34}$S, H$^{13}$CO$^+$, H$^{13}$CN, i.e. isotopic lines which are optically thin. While systematic errors are $\Delta V  <  1$ km/s, they find a limit of  $\Delta\mu/\mu < 1.4\, 10^{-6}$.

The NH$_3$ radio inversion lines were also used by Murphy et al (2008) in B0218, at z=0.7 who find no variations ($\Delta\mu/\mu < 1.8\, 10^{-6}$, corresponding to $\Delta v <$ 1.9km/s), and the same negative result was found with quasar Lyman band H$_2$ absorption lines, with improved spectral calibration and fitting all lines
in the Ly$\alpha$ forest, by King et al (2008):
$\Delta\mu/\mu = (2.6 \pm 3.0)\, 10^{-6}$ with redshift between 2 and 3.

While fitting the B0218 absorption lines, Murphy et al (2008) split the profiles in a large number of gaussian components, which could sometimes appear unrealistic. However, the optimal number of components minimizing the $\chi^2$ is 8.

\section {Main caveats with the radio molecular absorption method}
 Up to now, fortunately, the radio absorbers have provided
only upper limits, and have not too much suffered from these caveats:

1- The assumption that all absorbing components are at the same velocity, whatever the molecule could be wrong, and this might happen for several reasons, chemistry, excitation, temperature, density etc...

2- The lines could be at different optical depths, which might shift the resulting average velocity

3- Some lines have resolved hyperfine components (HCN, NH$_3$, ..) and the assumption of LTE for their excitation might fail

4- There could be completely different material in projection on the line of sight, with different origin and abundances, or different covering factor

5- At the various frequencies observed, the background continuum source has different sizes, and the absorbing medium is not the same 

6- The continuum source varies in both intensity and shape; and the compared lines have not been observed simultaneously

Some of these problems could be cured by selecting high density tracers, optically thin isotopes, lines at similar continuum frequencies, and taking care of the simultaneity of observations. In any case, for the systematic velocity offsets, or the different chemistry components, only large statistics will smooth the errors down.

\section{Advantages of the radio method}

The radio method has the potentiality to give the best constraints,
provided that more systems are detected, since:

(1) The rest frequencies are more precisely known than in the optical, and there is no problem of frequency calibration, at the precision level required

(2) The sensitivity of the line position to the variation of constants is much higher ($K_i$ larger by a factor 100!)

(3) The isotopic lines are observed separated, while in the optical there is a blend, with possible differential saturation

(4) Also the hyperfine structure is resolved, while in the optical there could be differential saturation, and velocity shifts

(5) The spectral resolution can be very high (10$^6$, heterodyne) and the observed lines are very narrow 

\section{Progress with ALMA}
  The radio studies suffer from the small numbers of sources, and the fact that they are all at low redshift. Fortunately, the sensitivity of ALMA will be able to detect many more continuum sources, to search for absorption lines, and at larger redshifts. ALMA will have a much wider bandwidth, allowing the search of absorption, even if not previously detected in the optical or HI-21cm. The redshift will be obtained directly in the millimeter.

 Up to now, there is about 50 systems detected in absorption in HI-21cm, at $z > 0.1$. Curran et al (2006, 2007, 2008) have undertaken a new search in front of 13 sources, and found only one new HI-21cm detection. With GMRT, Gupta et al (2009) have found intervening HI-21cm absorbers, with a higher detection rate:  9 new 21cm detections out of 35 MgII systems, at z$\sim$ 1.3.
 It is possible to predict the number of continuum sources that can be selected as targets for absorption searches with ALMA. The density of flat-spectrum quasars has been shown to follow the same curve as optical quasars, a curve peaking at $z\sim 2$, very similar to the star formation history
(Wall et al 2005). They are however still of significant density at $z \sim 3$.

\section{Conclusion and perspectives}

Due to caveats (velocity offsets, variability, optical depths, hyperfine ratios, etc)
a negative result is more easy to believe than a positive one!

There is many ways to improve the results: first obtaining more systems and more statistics, then improving the sensitivity and the signal-to-noise ratios, taking into account the simultaneity of observations for the lines to be compared.
The number of continuum sources that will be use as targets for absorption studies will be one or two orders of magnitude larger than now with ALMA, although the millimeter domain is not favoured here by the K-correction.
 
It could then be interesting to search rest-frame 3mm systems at centimeter wavelengths, with EVLA, SKA precursors and in the future with SKA.

\begin{figure*}[t!]
\centerline{
\resizebox{11cm}{!}{\includegraphics[angle=-90,clip=true]{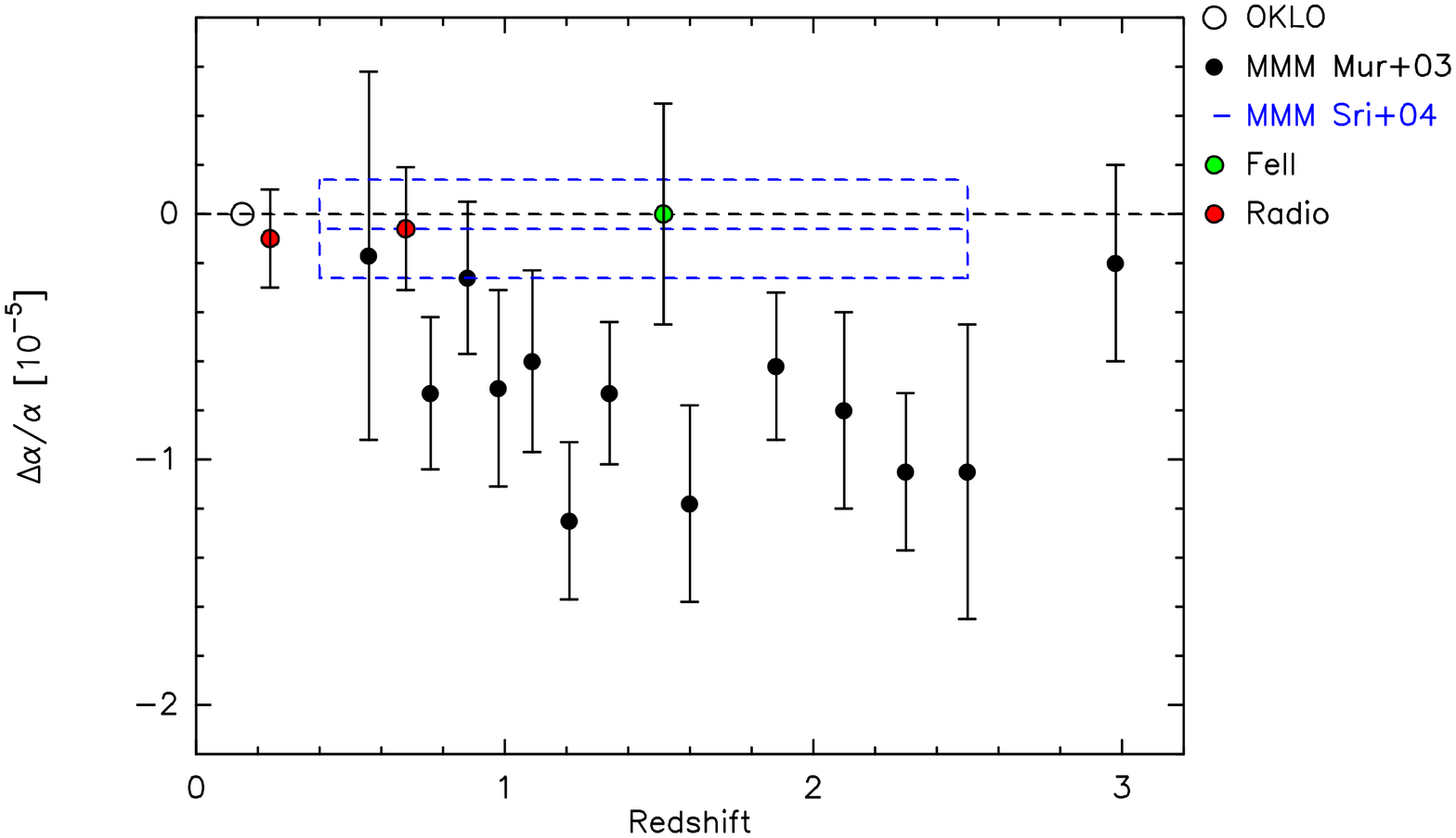}}
}
\centerline{
\resizebox{11cm}{!}{\includegraphics[angle=-90,clip=true]{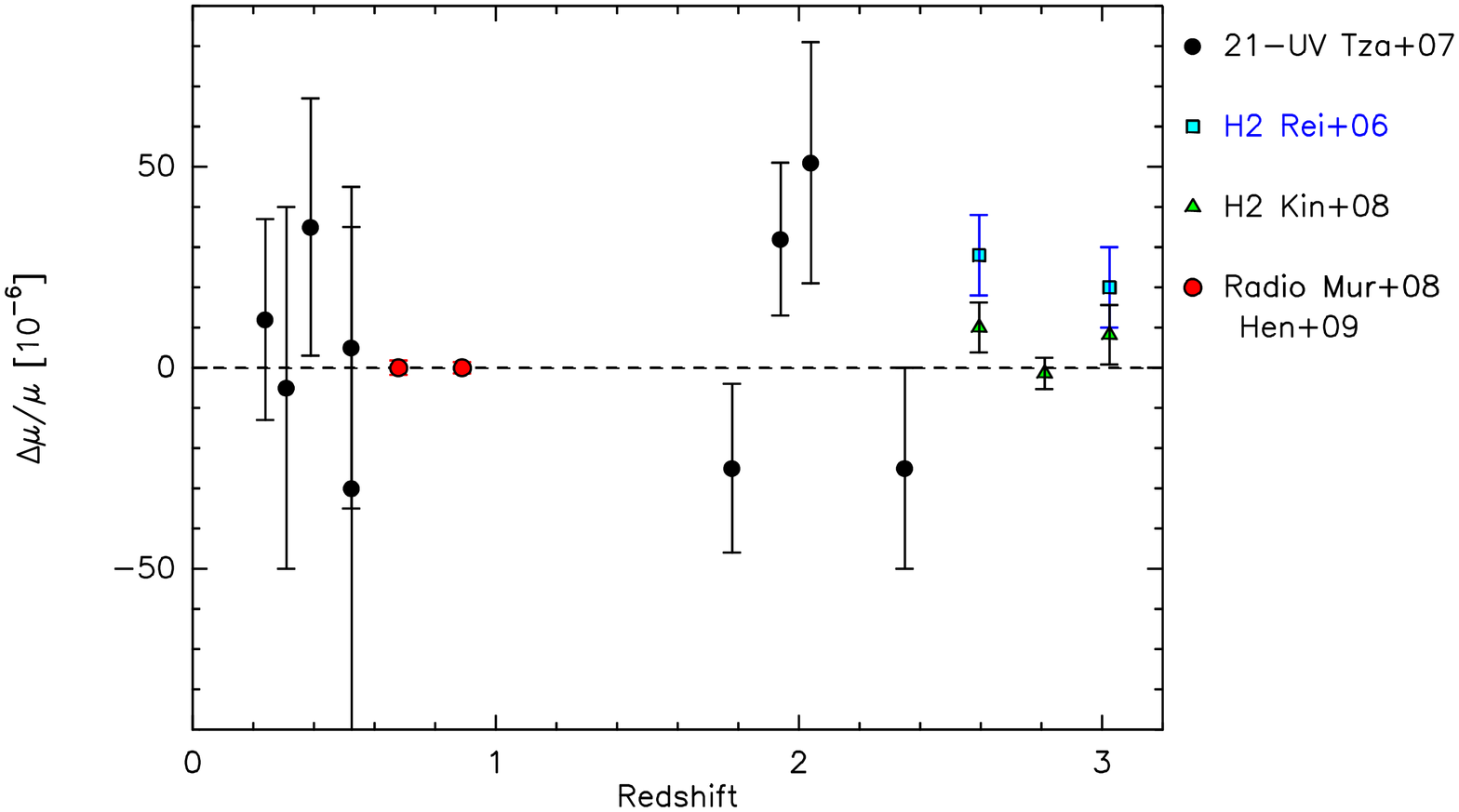}}
}
\caption{\footnotesize
Comparison between all observed constraints on variation for
$\alpha$ (top) and $\mu$ (down). The data in the $\alpha$ diagram are from OKLO (Damour \& Dyson 1996), MMM (Murphy et al 2003, Srianand et al 2004), FeII (Quast et al 2004) and radio (Murphy et al 2001). The data in the $\mu$ diagram are from 21-UV (Tzanavaris et al 2007), H$_2$ (Reinhold et al 2006, King et al 2008), and radio (Murphy et al 2008, Henkel et al 2009).
}
\label{alpha}
\end{figure*}

\begin{acknowledgements}
Many thanks to the organisers of this exciting joint discussion,
Paolo Molaro and Elisabeth Vangioni-Flam.
\end{acknowledgements}

\bibliographystyle{aa}

\end{document}